# Bifunctional Noble Metal-free Ternary Chalcogenide Electrocatalysts for Overall Water Splitting

Shantanu Singh[1,5], Ahamed Irshad[2], Germany Diaz De la Cruz[2], Boyang Zhao[1], Billal Zayat[2], Qiaowan Chang[3], Sri Narayan[2], Jayakanth Ravichandran[1,4,5]*

1: Mork Family Department of Chemical Engineering and Materials Science, University of Southern California, Los Angeles, California 90089, United States
2: Department of Chemistry, University of Southern California, Los Angeles, California 90089, United States
3: The Gene and Linda Voiland School of Chemical Engineering and Bioengineering, Washington State University, Pullman, Washington 99164, United States
4: Ming Hsieh Department of Electrical and Computer Engineering, University of Southern California, Los Angeles, California 90089, United States
5: Core Center of Excellence in Nano Imaging, University of Southern California, Los Angeles, California, 90089, United States

*Email: j.ravichandran@usc.edu

# Abstract:

Hydrogen has been identified as a clean, zero carbon, sustainable, and promising energy source for the future, and electrochemical water splitting for hydrogen production is an emission-free, efficient energy conversion technology. A major limitation of this approach is the unavailability of efficient, abundant, inexpensive catalysts, which prompts the need for new catalytic materials. Here, we report the synthesis and electrocatalytic properties of a novel transition metal-based ternary chalcogenide family, $LaMS_3$ (M = Mn, Fe, Co, Ni). Powder X-ray diffraction confirms the phase purity of these materials, while composition analysis using energy dispersive spectroscopy (EDS) confirms the presence of the stoichiometric ratio of elements in these compounds. X-ray photoelectron spectroscopy (XPS) and X-ray absorption spectroscopy (XAS) were used to study the chemical states on the surface and in bulk, respectively. These materials exhibit bifunctional catalytic activity towards the two half-reactions of the water-splitting process, with $LaNiS_3$ being the most active material for both hydrogen evolution reaction (HER) and oxygen evolution reaction (OER). The $LaMS_3$ compounds show long-term stability with negligible change in the overpotential at a constant current density of 10 mA $cm^{-2}$ over 18 hours of measurements. As compared to the corresponding ternary oxides, the $LaMS_3$ materials exhibit higher activity and significantly lower Tafel slopes. The ability to catalyze both half-reactions of water electrolysis makes these materials promising candidates for bifunctional catalysts and presents a new avenue to search for high-efficiency electrocatalysts for water splitting.

# Introduction:

Hydrogen is a promising energy carrier for future systems that can provide on-demand, emission-free energy at the point of use.[1,2] Electrochemical water splitting is a clean hydrogen production technique when renewable energy is used as the input. Water splits into hydrogen ($H_2$) at the cathode by HER and oxygen ($O_2$) at the anode by OER under an ideal thermodynamic potential of ~1.23 V,[3,4] but overpotentials are necessary to overcome the barriers for each of these reactions, which increases the associated cost of hydrogen production. Out of these two reactions, OER is kinetically sluggish for being a four-electron transfer reaction compared to HER, which is a two-electron transfer reaction.[5–7] Hence, the development of catalysts for OER is often considered more challenging compared to HER. Highly efficient electrocatalysts that can reduce the overpotentials for both half-reactions OER and HER are noted as bifunctional catalysts, and they can simplify the design of the electrochemical system for water splitting. Currently, Pt and Ru/ Ir-based materials are the state-of-the-art catalysts that exhibit high catalytic activities for HER and OER, respectively.[3,7,8] Their scalable implementation is limited by the high costs, earth abundance, and the production capacity of the precious metals, especially for OER. This has led to increased efforts to find catalysts for water splitting that are free of noble metals.[9–13]

First-row late transition metal (Mn, Fe, Ni, Co) based compounds have been widely investigated as alternates to the noble metal-based catalysts for their earth abundance and multivalent nature of these elements arising from their $d$-orbital character.[14–16] A class of materials that has been widely explored in this direction are transition metal oxides. Sustained research on optimizing the composition and control over the microstructure of these materials has led to improvements in their performance.[5,17–20] While these materials show good activity towards OER, they typically suffer from issues such as instability and low conductivity.[21–23] Binary sulfides, selenides, and phosphides have also appeared as candidates for water splitting.[24–29] Although the performance of these materials is promising, the compositional space available to optimize the balance between activity and stability in these compounds remains limited. To address these limitations in the chemical design phase space, we explore ternary sulfides as candidates for electrocatalysts in water splitting. While ternary chalcogenides based on early transition metals, particularly Ti and Zr, have emerged as a family of electronic materials with potential applications in photovoltaics[30–32] and infrared optics,[33–35] our recent study on their electrocatalytic properties revealed large overpotentials and poor efficiency of water splitting.[36] As has been seen in oxides,[18] we presume that ternary chalcogenides based on late transition metals may have more attractive electrocatalytic properties. Among such ternary chalcogenides, $LaMS_3$ (M= Mn, Fe, Co, Ni) powders have been reported to crystallize in the hexagonal *P*6$_3$ space group at temperatures at or above 1100°C from the sulfurization of ternary oxide powders.[37,38] These studies report electrical and magnetic properties of these materials with semiconducting or metallic properties, which show that such chalcogenides have higher conductivity than equivalent oxides and likely better electrocatalytic properties.

In this article, we report the synthesis of $LaMS_3$, where M = (Mn, Fe, Ni, Co), by sulfurization of the corresponding oxide powders in a stream of Argon and Carbon disulfide at 1100-1200°C and for the first time, their electrocatalytic properties for water splitting. We verified the phase purity and chemical composition of these materials using powder X-ray diffraction measurements and EDS. Bulk and surface chemical states were analyzed using XAS and XPS measurements, respectively. We made these powders into suspensions and measured their catalytic activity toward water-splitting reactions by preparing electrodes *via* drop casting. $LaMS_3$

compounds demonstrate appreciable activity towards both HER and OER in alkaline medium and towards HER in acidic medium. Among these compounds, LaNiS$_3$ exhibits the highest activity towards both HER and OER, with low overpotentials and Tafel slope values and excellent stability over several hours of measurement, exhibiting negligible change in overpotential over time. We also report the comparison of catalytic performance of the chalcogenides with corresponding oxides, and LaMS$_3$ compounds show lower overpotential and lower Tafel slope values for both HER and OER.

## Results and Discussions:

We synthesized polycrystalline powders of $LaMS_3$ by annealing $LaMO_3$ powders in a stream of Argon saturated with $CS_2$ at temperatures ranging from 600 to 1200°C for annealing times ranging from 4 to 30 hours (SI, Figure S1). Based on our observations, powders annealed at temperatures below 1100°C were partially sulfurized with either unreacted oxide left in the product or lanthanum oxysulfide ($La_{10}OS_{14}$) as the major product. On the other hand, samples annealed at temperatures above 1100°C resulted in pure phases of $LaMS_3$ in the final product. Diffraction patterns obtained from room temperature powder diffraction are shown in Figure 1(a). The diffraction patterns for all the materials are in agreement with the patterns reported in the past.[37] All $LaMS_3$ compounds reported are isostructural and crystallize in a hexagonal $P6_3$ symmetry (as shown in Figure 1(b)). The lattice parameters for these materials are reported in Table 1. Chemical compositions were determined for these materials using energy dispersive spectroscopy (EDS) to establish the atomic ratio between different elements. The result obtained was then calibrated using $LaMO_3$ as a reference for the La: M ratio. The atomic ratios obtained after the corrections match well with the expected stoichiometric ratio of 1:1:3 for La: M: S (SI, Figure S2).

To analyze the chemical states of $LaNiS_3$ and $LaCoS_3$ on the surface and in the bulk, we utilized X-ray photoelectron spectroscopy (XPS) and X-ray absorption near-edge structure (XANES), respectively (Figure 2). XPS analysis demonstrates Ni-S bond formation on the surface of the $LaNiS_3$ samples, with Ni $2p$ peaks observed at 870.3 eV (Ni $2p_{1/2}$) and 853.2 eV (Ni $2p_{3/2}$), and S 2p peaks at 162.9 eV (S $2p_{1/2}$) and 161.7 eV (S $2p_{3/2}$), are characteristic of the Ni-S bond.[39,40] In addition to this, a low-intensity peak centered around 857 eV is observed, indicating the formation of sulfate ($SO_4^{2-}$) on the surface.[41,42] XANES analysis further confirms these findings by the observation that the white line intensity of the Ni *K*-edge spectrum for LaNiS$_3$ falls between the NiS and Ni$_2$S$_3$ standards, as shown in Figure 2(c). This suggests the presence of a Ni-S coordination environment, which aligns with the results obtained from XPS and bulk-sensitive X-ray diffraction results. In contrast, LaCoS$_3$ presents a distinct chemical state on its surface, as revealed by XPS analysis. The Co 2p XPS spectrum identifies both Co-O and Co-S formation. For instance, the Co $2p_{3/2}$ component is deconvolved into three peaks, including the Co−S peak at 778.4 eV, the Co−O peak at 779.6 eV, and a satellite peak at 781.8 eV.[43–45] The S 2p spectrum, displaying binding energies at 161.7 and 162.9 eV, confirms the typical presence of $S^{2-}$, consistent with expectations for Co-S interactions. However, the XANES Co K-edge spectrum of $LaCoS_3$ is nearly identical to that of CoS, showing a similar pre-edge intensity from the Co 1s to 3d transitions and a comparable white line intensity, which indicates the presence of Co-S as the dominant species. The comparison of the XPS and XANES results highlights the difference between the surface and bulk structures of $LaCoS_3$; while the surface of $LaCoS_3$ exhibits partial oxidation, its bulk retains the Co-S coordination environment.

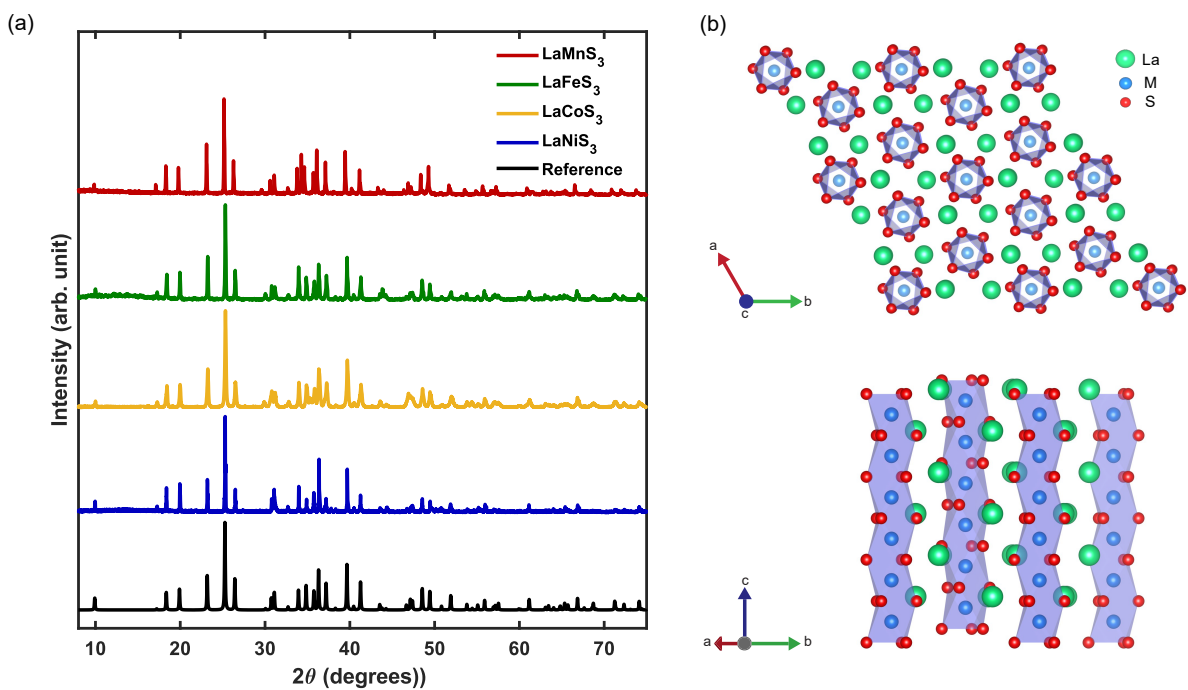

**Figure 1: (a)** Powder X-ray diffraction pattern with reference for LaMnS$_3$, LaFeS$_3$, LaCoS$_3$, and LaNiS$_3$. **(b)** Schematic crystal structure of LaNiS$_3$ viewed along the c-axis, showing hexagonal symmetry (top) and with c-axis in plane, showing NiS$_6$ chains parallel to the c-axis (bottom).

**Table 1:** Lattice parameters for LaMS$_3$ materials.

| Material | Space Group | Lattice Parameters | |
| --- | --- | --- | --- |
| | | a (Å) | c (Å) |
| LaMnS$_3$ | *P*6$_3$ | 10.3152 | 5.7526 |
| LaFeS$_3$ | | 10.3112 | 5.7566 |
| LaCoS$_3$ | | 10.2988 | 5.7518 |
| LaNiS$_3$ | | 10.2841 | 5.7573 |

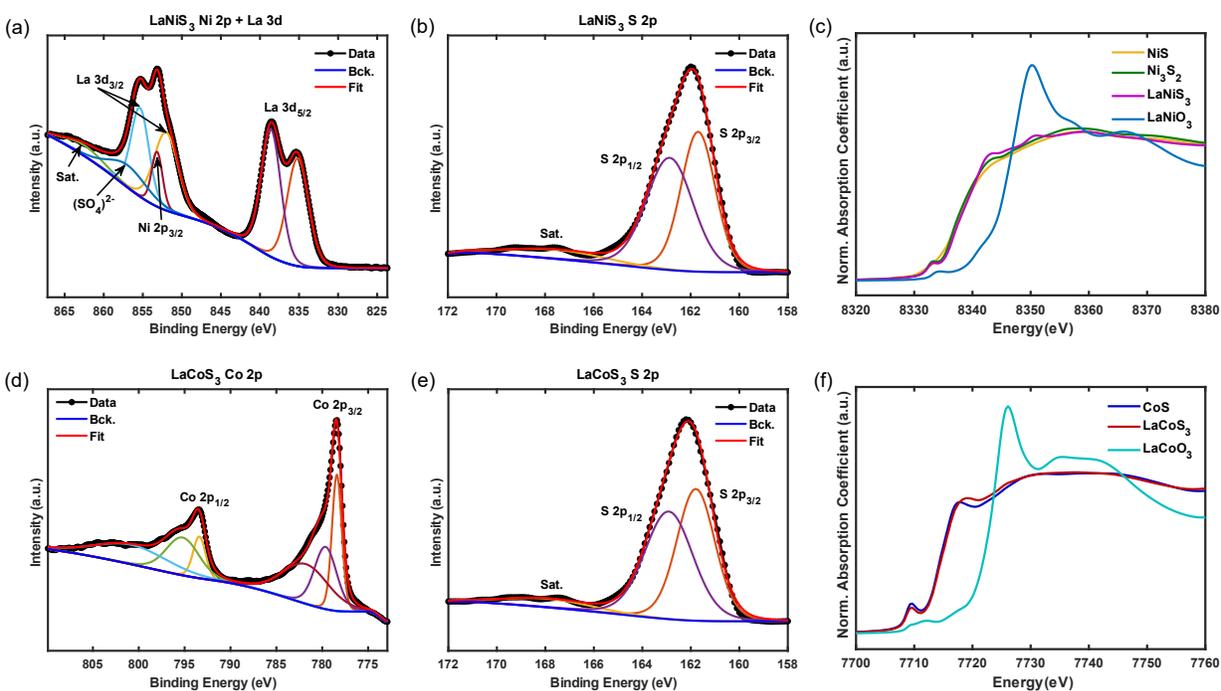

**Figure 2:** XPS and XAS spectra for $LaNiS_3$ and $LaCoS_3$. High resolution XPS spectra for **(a)** Ni 2p and **(b)** S 2p peaks, and **(c)** Ni-K edge XANES spectra for $LaNiS_3$; and high resolution XPS spectra for **(d)** Co 2p and **(e)** S 2p peaks, and **(f)** Co-K edge XANES spectra for $LaCoS_3$. Spectra for relevant binary sulfides and ternary oxides are shown for comparison.

Testing for the catalytic activity of LaMS$_3$, we first characterized the OER electrocatalytic performance of these materials. A standard three-electrode system with nickel foil counter electrode and 1 M KOH electrolyte was used to evaluate the OER electrocatalytic performance of the LaMS$_3$ deposits on the electrodes achieved from suspensions, and the linear sweep voltammograms (LSVs) measured at 1 mV s$^{-1}$ and 2000 rpm are shown in Figure 3(a). The slow scan rate ensures steady state condition, and electrode rotation avoids mass transfer limitations and removes gas bubbles from the electrode surface. Based on these measurements, it is evident that the LaNiS$_3$ and LaCoS$_3$ powders show good catalytic activity towards OER.

The LaNiS$_3$ sample shows the highest activity with a low onset potential of 1.49 V, and 10 mA cm$^{-2}$ current density is achieved at an overpotential ($\eta_{10}$) of 373 mV (overpotential for OER = E$_{applied}$ (vs. RHE) – 1.23). Similarly, the cobalt-containing sample (LaCoS$_3$) starts to evolve oxygen at around the same potential; however, 50 mV higher overpotential than LaNiS$_3$ is required to reach 10 mA cm$^{-2}$, suggesting a slightly lower catalytic activity. No significant gas evolution is observed for both LaFeS$_3$ and LaMnS$_3$ samples, even at 1.72 V. Nonetheless, LaFeS$_3$ is found to be slightly more active than LaMnS$_3$. For instance, at 500 mV overpotential, LaFeS$_3$ yields 1.9 mA cm$^{-2}$ current, whereas LaMnS$_3$ shows only 0.5 mA cm$^{-2}$. Thus, a periodic trend in the OER activity as LaNiS$_3$ > LaCoS$_3$ >> LaFeS$_3$ > LaMnS$_3$ could be noticed.

We also investigated whether the high activity of LaNiS$_3$ and the activity trend originate from the intrinsic property of the materials or exclusively from the differences in their electrochemical surface areas (ECSAs). The ECSAs and roughness factors obtained from the scan rate dependence of cyclic voltammetry current in the non-Faradaic region (SI, Figure S4) are similar for all the catalysts around 0.5 cm² and 2.5, respectively (SI, table S2). Hence, the higher activity of LaNiS$_3$ cannot be attributed to the enhanced number of surface-active sites.

**Table 2:** The Kobussen OER Pathway

|  | Rate Determining Step | $\partial \eta / \partial \ln i$ | |
| --- | --- | --- | --- |
|  |  | Low OH⁻ Coverage | High OH⁻ Coverage |
| Step 1 | $M + OH^- \rightarrow MOH + e^-$ | 2RT/F | -- |
| Step 2 | $MOH + OH^- \rightarrow MO + H_2O + e^-$ | 2RT/3F | 2RT/3F |
| Step 3 | $MO + OH^- \rightarrow MO_2H^-$ | RT/2F | ∞ |
| Step 4 | $MO_2H^- + OH^- \rightarrow MO_2^- + H_2O + e^-$ | 2RT/5F | 2RT/F |
| Step 5 | $MO_2^- \rightarrow M + O_2 + e^-$ | 2RT/7F | ∞ |

*Note: 2.303 RT/F = 0.059 V

To further analyze the reaction kinetics and the OER mechanism, Tafel plots are derived from the polarization curves, as shown in Figure 3(b). The Tafel slope is a key parameter that mainly depends on the reaction mechanism and the nature of the rate-determining step (RDS).[46] A lower value of the Tafel slope indicates a faster electrochemical reaction. Here, LaNiS$_3$ shows a Tafel slope value of 48 mV dec⁻¹ which is much lower than the values for LaCoS$_3$ (66 mV dec⁻¹), LaFeS$_3$ (194 mV dec⁻¹), and LaMnS$_3$ (345 mV dec⁻¹), indicating that LaNiS$_3$ is the best-suited material among the series for OER, both in terms of low overpotential and small Tafel slope. In general, the OER mechanism is broadly classified into the conventional adsorbate evolution mechanism (AEM) and the lattice oxygen mechanism (LOM), depending on the origin of oxygen atoms.[47] In the case of AEM, all the oxygen atoms in the evolved oxygen come from the adsorbed oxygen species (OH$^-$/H$_2$O), whereas in LOM, all or partial oxygen atoms of the oxygen gas are from the lattice oxygen atoms of the catalyst material. Although LOM, particularly in perovskite oxides, can provide higher OER activity, the involvement of lattice oxygen could lead to inherent instability of the catalyst.[48] In the absence of oxygen atoms in the lattice, AEM is the most probable mechanism for the sulfide catalysts tested here. Based on early reports from Bockris *et al.* on the adsorbate-based OER mechanism on perovskites,[49] the Kobussen Pathway (Table 2) seems to be the most viable for LaNiS$_3$, with the RDS being step 2, since regardless of OH$^-$ coverage, the predicted Tafel slope of 2.303*2RT/3F (~40 mV dec$^{-1}$) is close to the experimental value (48 mV dec$^{-1}$). The results are also consistent with our previously published work on nickel-based OER catalysts in alkaline electrolyte.[50] The OER activity of the catalyst for this type of mechanism is highly dependent on the adsorption strength of the intermediates. However, the adsorption energies of the various intermediates are linearly correlated, which sets the lower limit of the OER overpotential at 370 mV.[51] In the case of LaNiS$_3$, the overpotential at 10 mA cm$^{-2}$ is 373 mV, which is close to the theoretical limit. Further improving the OER activity may require breaking such a scaling relationship, which provides the clue for the next steps in the design of better-performing catalysts.

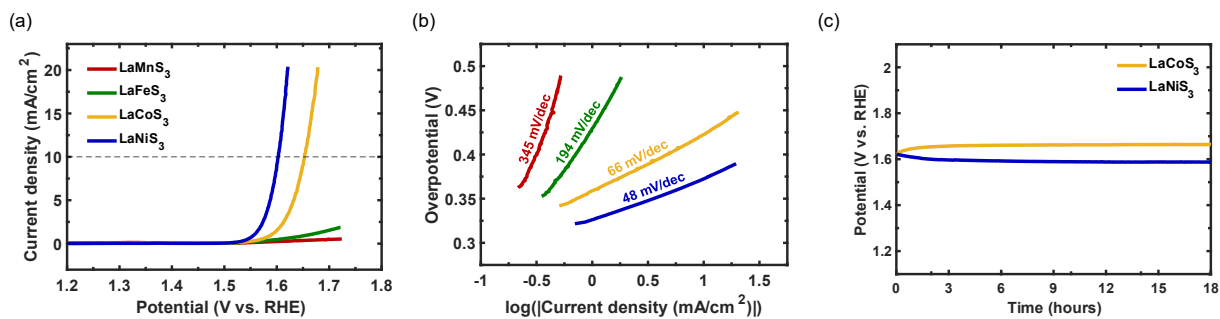

**Figure 3:** OER activity tests in 1 M KOH solution. **(a)** Linear sweep voltammograms (LSVs), **(b)** Tafel plots in the linear region, and **(c)** stability test at 10 mA cm$^{-2}$ for 18 hours.

The long-term operation stability of the catalyst is crucial for their practical application. Under the highly oxidizing condition of OER, most of the catalysts will be corroded, leading to a significant increase in the overpotential. Hence, we tested the stability of the best-performing $LaNiS_3$ and $LaCoS_3$ catalysts with chronopotentiometry over an 18-hour duration at a constant current density of 10 mA cm$^{-2}$. As shown in Figure 3c, both $LaCoS_3$ and $LaNiS_3$ exhibit high stability with no appreciable changes in the required potential during continuous operation. $LaNiS_3$ shows slight enhancement in the activity at the end of 18 hours, probably due to improved surface area caused by vigorous gas evolution. The overall change in the overpotential was ~1 mV h$^{-1}$ for both $LaCoS_3$ and $LaNiS_3$ during continuous oxygen evolution at 10 mA cm$^{-2}$ for 18 hours.

HER is usually studied in an acidic medium due to the abundant supply of the reactant species (H$^+$ or $H_3O^+$) and relatively fast kinetics. Hence, we first examined the HER catalytic activity and stability of the materials in the $LaMS_3$ series in 0.5 M $H_2SO_4$. The linear sweep voltammograms measured at 1 mV s$^{-1}$ scan rate and 2000 rpm electrode rotation for all the sulfides are shown in Figure 4(a); the required overpotentials (overpotential for HER = |E$_{applied}$ (vs. RHE) – 0|) for 10 mA cm$^{-2}$ HER current densities ($\eta_{10}$) are 340, 375, 558, and 655 mV for $LaNiS_3$, $LaCoS_3$, $LaFeS_3$, and $LaMnS_3$, respectively. Although the onset potential of hydrogen evolution is similar around -0.21 V for both $LaCoS_3$ and $LaNiS_3$ catalysts, $LaNiS_3$ shows much better performance at a more negative potential (E < -0.30 V) due to its lower Tafel slope compared to $LaCoS_3$ (as discussed below). Overall, a periodic change in the HER activity of $LaMS_3$ with M-site cation could be observed (i.e., Ni > Co >> Fe > Mn), which is similar to that in OER and is maintained even after the normalization to ECSAs. Thus, $LaNiS_3$ shows the highest OER and HER intrinsic activity among all the catalysts tested here. According to the Sabatier principle, the binding energy of the catalyst with the reactant (H$^+$ in this case) should be optimum for high catalytic activity.[52] In perovskite oxides of the general form AMO$_3$, the transition metal ions located at the M-sites are often considered the active sites for the HER.[53] Therefore, the periodic change in the activity could be attributed to a gradual variation in the binding strength of the H$^+$ ion with transition metal cation at the M-site. This is also consistent with the Tafel slope analysis discussed later, which suggests weak adsorption of H$^+$ as the rate-limiting process. It is therefore anticipated that further enhancement in the activity could be achieved by tuning the binding strength using partial doping at the A-site or M-site, straining the surface, or creating more anion vacancies. In addition to the highest activity, LaNiS$_3$ exhibits the least Tafel slope of 79 mV dec$^{-1}$ and high stability during 18 hours of electrolysis at 10 mA cm$^{-2}$ with only ~3 mV h$^{-1}$ increase in the overpotential (Figures 4(b) and 4(c)). Other catalysts are also stable, with a potential change in the range of 1-5 mV h$^{-1}$ during continuous operation at 10 mA cm$^{-2}$ for 18 hours. The Tafel slope values are 147, 116, and 117 mV dec$^{-1}$ for LaCoS$_3$, LaFeS$_3$, and LaMnS$_3$, respectively.

In general, HER in acidic medium follows,

Volmer: $\quad H_3O^+ + + 1e^- + M \rightleftharpoons M - H + H_2O$

Tafel: $\quad 2M - H \rightleftharpoons 2M + H_2$

Heyrovsky: $\quad M - H + H_3O^+ + + 1e^- \rightleftharpoons M + H_2O + H_2$

Here, "M" refers to an empty surface site of the catalyst, and the "M-H" refers to the absorbed hydrogen atom. Theoretically (at 25°C and charge transfer coefficient, α = 0.5), if the Volmer step (electrosorption of H$^+$) is the RDS, a Tafel slope of 120 mV dec$^{-1}$ is expected, whereas 30 and 40 mV dec$^{-1}$ are predicted for Tafel and Heyrovsky steps as the RDSs. The measured Tafel slopes in all the catalysts except for LaNiS$_3$ are close to 120 mV dec$^{-1}$, suggesting the Volmer step as the RDS.

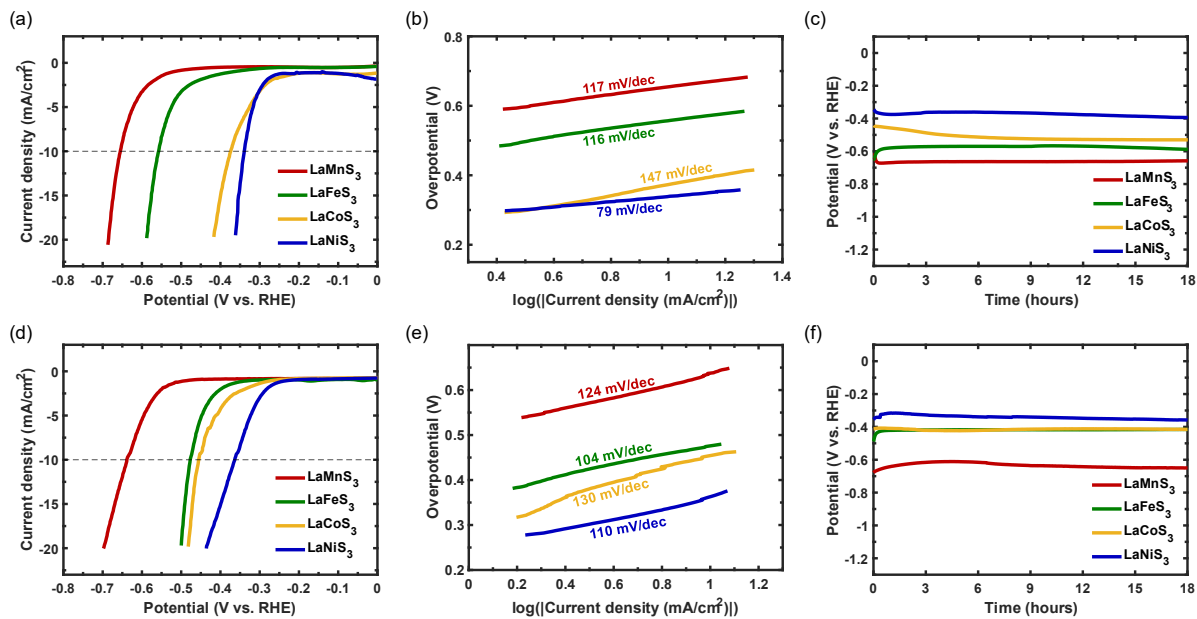

**Figure 4:** HER activity tests in 0.5 M $H_2SO_4$ and 1 M KOH. **(a)** and **(d)** Linear sweep voltammograms, **(b)** and **(e)** Tafel plots in the linear region, and **(c)** and **(f)** stability test at 10 mA cm$^{-2}$ for 18 hours.

Similar values have been reported for the weakly proton binding transition metals such as Au or Ag, which shows the Volmer step as the RDS, followed by either electrochemical Heyrovsky or chemical Tafel steps.[54] The Tafel step occurs only when the distance between the surface adsorbed hydrogen is shorter than the Van der Waals radius of the two adsorbed hydrogen atoms. Conversely, the Volmer-Heyrovsky mechanism dominates when the active sites are diluted, and the distance between two adsorbed hydrogen is larger than the Van der Waals radius, which is more likely to occur in the present case.[55] Nevertheless, the Tafel slope obtained in the case of $LaNiS_3$ doesn't match with any of the predicted values, but it is comparable with the values reported for Ni-Co-S,[56] Co−S,[57] Ni−S,[40] and $MoS_3$.[58]

We also tested the HER activity in alkaline electrolytes. Compared to acidic electrolytes, water splitting in alkaline media can operate with non-noble metals-based catalysts and provide better stability. However, the activity of HER in alkaline media, even for noble metals, is usually 2-3 orders of magnitude lower than in acids.

In the case of HER in acidic solution, as discussed above, the proton adsorption binding energy plays a key role. However, in alkaline conditions, the activity is controlled by the balance of three major criteria: (i) water dissociation to produce free H$^+$ in solution, (ii) the adsorption of H$^+$ on the active sites, and (iii) poisoning of active sites by OH$^-$ adsorption.[59] HER activity is more sluggish in alkaline medium primarily due to additional energy needed to break the strong covalent H-O-H bond in the Volmer step rather than the weak dative bond of H$_3$O$^+$ ions in acids.

Volmer step (alkaline medium): 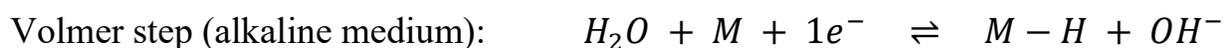

$$H_2O + M + 1e^- \rightleftharpoons M-H + OH^-$$

Hence, the electrocatalysts that can dissociate water efficiently and have optimum binding strength with hydrogen will exhibit high HER activity in alkaline condition. Interestingly, the HER activity of LaNiS$_3$ in alkaline medium is comparable to that in acid with an overpotential requirement of 362 mV at 10 mA cm$^{-2}$, whereas LaCoS$_3$, LaFeS$_3$, and LaMnS$_3$ require 450, 478, and 640 mV, respectively at the same current density (Figure 4(d)). The Tafel slope values are 110, 130, 104, and 124 mV dec$^{-1}$ for LaNiS$_3$, LaCoS$_3$, LaFeS$_3$, and LaMnS$_3$, consistent with the Volmer step as the RDS (Figure 4(e)). Since water dissociation and subsequent formation of adsorbed hydrogen is the RDS in alkaline medium, the excellent HER activity of LaNiS$_3$ indicates its ability to dissociate water efficiently. This is also supported by the reports from Markovic *et al.* on improving the HER activity of metals in alkaline electrolyte by modifying the surface with Ni(OH)$_2$ water dissociation catalyst.[60] In addition, LaNiS$_3$ exhibited remarkable stability with only 0.2 mV h$^{-1}$ change in the overpotential during 18 hours of continuous operation at a current density of 10 mA cm$^{-2}$ (Figure 4(f)). LaCoS$_3$ and LaMnS$_3$ also show a negligible increase in overpotential of 0.35 mV h$^{-1}$ and 1.7 mV h$^{-1}$. Interestingly, LaFeS$_3$ shows some activation, leading to a decrease in the required overpotential at 10 mA cm$^{-2}$ by 0.08 V in 18 hours.

To further illustrate the advantages of sulfurization, we compared the OER performance of LaCoS$_3$ and LaNiS$_3$ with their parent oxides (i.e., LaCoO$_3$ and LaNiO$_3$) under identical conditions in 1 M KOH. Figure 5(a) suggests that sulfides are more active towards OER than oxides with LaCoO$_3$ and LaNiO$_3$ requiring 53 and 15 mV higher overpotentials at 10 mA cm$^{-2}$ than LaCoS$_3$ and LaNiS$_3$, respectively. In addition, LaCoO$_3$ and LaNiO$_3$ show relatively higher Tafel slopes of 90 mV dec$^{-1}$ and 79 mV dec$^{-1}$, respectively. These results clearly suggest that sulfides are better OER catalysts than their corresponding oxides, and among all the catalysts tested here, LaNiS$_3$ requires the least overpotential at 10 mA cm$^{-2}$ current density and exhibits the smallest Tafel slope and excellent durability in 1 M KOH. The OER activity of LaNiS$_3$ is also comparable to those of many reported oxides and sulfides.[9,10]

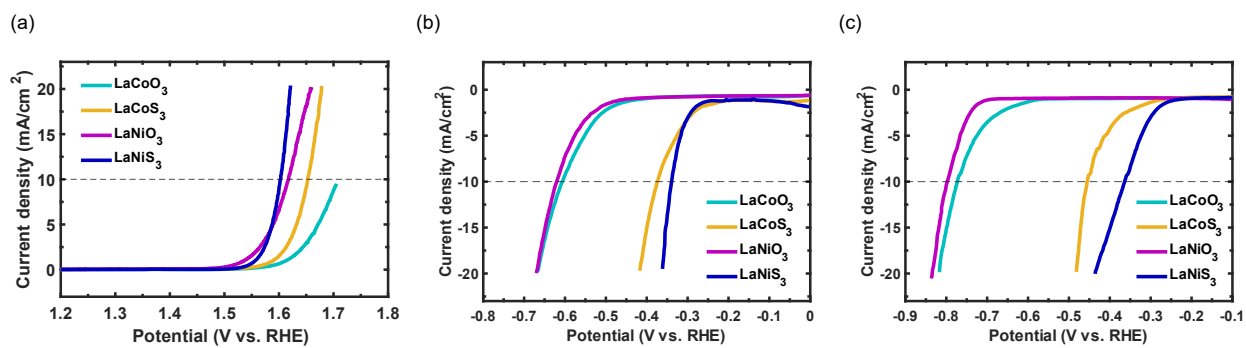

**Figure 5:** Linear sweep voltammograms (LSVs) comparing selected sulfides with corresponding oxides. **(a)** LSVs comparing OER performance in 1 M KOH, and **(b)** and **(c)** LSVs comparing HER performance in 0.5 M H$_2$SO$_4$ and 1 M KOH, respectively.

Similarly, we also compared the HER performance of sulfides with oxides in both alkali and acidic media. The oxide catalysts, LaCoO$_3$ and LaNiO$_3$, require more than 250 mV higher overpotential than their corresponding sulfides, LaCoS$_3$ and LaNiS$_3$ at 10 mA cm$^{-2}$, both in 0.5 M H$_2$SO$_4$ and 1 M KOH solutions (Figures 5(b) and 5(c)), which indicates the advantageous role of sulfurization towards HER in a wide range of pH. It is also observed that the difference in activity for the same catalyst between acidic and alkaline media is higher in the case of oxides than sulfides. For instance, LaNiS$_3$ exhibits overpotential values of 340 and 362 mV at 10 mA cm$^{-2}$ in acidic and alkaline media, implying only a 22 mV increase in the overpotential on switching the electrolytes. On the other hand, LaNiO$_3$ shows a substantial increase in overpotential of around 170 mV at 10 mA cm$^{-2}$ when the electrolyte is changed from acidic to alkaline. This again can be attributed to the higher efficiency of the sulfides to dissociate water in the Volmer step of alkaline HER than oxides. Furthermore, sulfides are electronically more conducting than oxides.[37] Since carbon is thermodynamically stable under the hydrogen evolution potential, HER testing of catalysts was done by adding 20 wt% SuperP carbon. Although the high surface area of carbon ensures electrical connection between individual particles, carbon plays a relatively less significant role in the case of sulfides compared to oxides. For example, when HER testing was performed without carbon, the required overpotential for 10 mA cm$^{-2}$ is increased by 145 mV for LaNiO$_3$, whereas the increase is only 48 mV for LaNiS$_3$ (SI, Figure S6). This suggests that, unlike the case of oxides, the addition of carbon may not be crucial for the sulfides, especially when tested as nanoparticles or thin films.

## Conclusion:

In conclusion, we have introduced a new family of materials, complex ternary transition metal chalcogenides, as catalysts for water splitting. We synthesized a series in this family of complex ternary transition metal chalcogenides (LaMS$_3$) using CS$_2$ sulfurization of corresponding ternary oxides at high temperatures. These materials, when tested as catalysts for the two half-reactions of water splitting, show good activity towards OER in alkaline medium and HER in both alkaline and acidic media. Among all the materials in the series, LaNiS$_3$ exhibits the best activity, with low overpotentials and low Tafel slope values, while exhibiting excellent long-term stability for both HER and OER. When compared to corresponding oxides, these materials demonstrate better activity in terms of lower overpotentials and smaller Tafel slopes. This study opens up an opportunity to explore the wider compositional space offered by complex chalcogenides for electrocatalysts and shows a key step towards achieving optimum activity for water splitting using noble metal-free sulfide catalysts.

## Methods:

**Synthesis of LaMS$_3$ powders:** We synthesized the polycrystalline powders of LaMS$_3$ using the following approach. First, precursor polycrystalline powders of LaMO$_3$ were synthesized using the standard synthesis methods reported for the corresponding ternary oxides. We used high-temperature solid-state reaction method to synthesize LaCoO$_3$ and LaFeO$_3$. First, La$_2$O$_3$ was heated to 600°C for 10 hours before use to get rid of any hydroxide impurities. Then, binary oxides La$_2$O$_3$ (Alfa Aesar, 99.99%) and Co$_3$O$_4$ (Sigma Alrich, 99.99%), and Fe$_2$O$_3$ (Alfa Aesar, 99.9%) in corresponding molar ratios were mixed together in an agate mortar and pestle till a uniform powder mixture was attained. After that, the powder mixtures were packed in an alumina crucible and heated to 1300°C with a ramp rate of 100°C h$^{-1}$, held for 48 hours, and allowed to cool down naturally by turning off the furnace. In case of LaNiO$_3$ and LaMnO$_3$, citrate-nitrate sol gel synthesis was used where La(NO)$_3$.6H$_2$O (Alfa Aesar, 99.99%), Ni(NO)$_3$.6H$_2$O (Alfa Aesar, 99.9985%), and Mn(NO)$_3$.xH$_2$O (ProChem Inc., 99.999%) were used as the precursors. Nitrates of relevant precursors in appropriate molar ratios were dissolved in deionized water, along with citric acid monohydrate (C$_6$H$_8$O$_7$.H$_2$O) (Sigma Aldrich, 99.0%) such that the molar ratio of metal nitrates to citric acid is 1:2. The solution thus obtained was dried at 120°C for 24 hours on a hot plate, followed by annealing at 700°C for 4 hours, with a heating/ cooling rate of 300°C h$^{-1}$. Next, LaMS$_3$ powders were synthesized by annealing the LaMO$_3$ powders in a carbon disulfide annealing furnace. LaMO$_3$ powders were loosely packed in an alumina crucible and loaded in a tube furnace of 1" dia. Samples were heated to 1100-1200°C, with a ramp rate of 10 °C min$^{-1}$, and held for 4-30 hours, and Argon was bubbled through CS$_2$ (Alfa Aesar, 99.9%), with the flow rate being 10-15 sccm. After the hold time, the furnace was turned off, and the samples were allowed to cool down naturally.

**Powder Characterizations:** Structural characterization using powder diffraction was carried out using a Bruker D8 Advance diffractometer using Cu K$\alpha$ wavelength at room temperature. Energy dispersive X-ray spectroscopy measurements were done in a Nova NanoSEM 450 Field Emission Scanning Electron Microscope, equipped with an OXFORD Ultim Max detector using an acceleration voltage of 20kV. The measurements were made using 6mm diameter pellets made from LaMS$_3$ powders using YLJ-100E 100T Electric Hydraulic Press from MTI. X-ray photoelectron spectroscopy (XPS) measurements were performed using a Kratos Axis Ultra DLD - X-ray Photoelectron Spectrometer, equipped with a magnetic immersion lens and charge neutralization system with the spherical mirror and concentric hemispherical analyzers. Powder samples were pressed into 6mm pellets for measurement. The monochromatic X-ray source used was the Al K$\alpha$ line at 1486.6 eV, and a pressure of <4x10$^{-8}$ Torr was maintained. Survey scans were carried out to collect the low-resolution spectra in the energy range of 0-1200 eV, with a step size of 1 eV, and a pass energy of 160 eV. High-resolution spectra were collected with a step size of 0.1 eV in the specific energy region of interest, with a pass energy of 40 eV. Data analysis was done using CasaXPS software. X-ray absorption near edge spectroscopy (XANES) measurements were done at the Stanford Synchrotron Radiation Lightsource (SSRL) at the 4-1 beamline. Samples were prepared by sandwiching the powders of relevant materials between two layers of Kapton tape. Measurements were done at liquid nitrogen temperature at the transition metal *K*-edge in fluorescence mode using a Lytle-type fluorescence-yield ion chamber detector. Energy for each material was calibrated using the respective standard metallic foil sample. Data processing was done using the Athena software of the Demeter package.

**Electrochemical Characterizations:** All electrochemical measurements were performed using a VersaSTAT multi-channel potentiostat (AMETEK Scientific Instruments) in a three-electrode configuration with 1 M KOH or 0.5 M $H_2SO_4$ aqueous solution as the electrolyte. High-purity water (18.2 MOhm cm) was used to prepare the electrolyte solutions. The catalysts loaded gold RDE (5 mm diameter) and glassy carbon RDE (5 mm diameter) were used as the working electrodes for the OER and HER, respectively. A graphite rod counter electrode and $Hg/HgSO_4$ (MSE in 0.5 M $H_2SO_4$) reference electrode were used in acidic medium, whereas nickel foil counter and Hg/HgO (MMO in 1 M KOH) reference electrodes were used for the studies in alkaline medium. Prior to the catalyst ink preparation, all the materials were ground gently using a mortar and pestle. Then, 10 mg of the catalyst powder was added to 1 mL of DMSO solvent containing 10 µL of 5 wt % Nafion binder. In the case of HER studies, 20 wt % SuperP carbon (Alfa Aesar) was also added unless otherwise noted. The mixture was vortex mixed followed by sonication for 2 hours at 30 min intervals to get a homogeneous catalyst ink. 20/ 30 µL of the ink was drop-cast onto the surface of a polished RDE and dried at 80°C for HER/ OER. The resulting catalyst loading was roughly 1 and 1.5 mg cm$^{-2}$ for HER and OER respectively. The HER and OER activity were measured using linear sweep voltammetry (LSV) at 1 mV s$^{-1}$ scan rate and electrode rotation of 2000 rpm. Stability measurements were carried out at 10 mA cm$^{-2}$ for 18 hours at 2000 rpm. Electrochemical surface area (ECSA) measurements were done using cyclic voltammetry in the non-Faradaic region at 25, 50, 100, 200, and 400 mV s$^{-1}$ scan rates. Electrochemical impedance spectroscopy (EIS) measurements were performed at the desired dc potential with a sinusoidal excitation voltage of 5-10 mV in the frequency range of 100 kHz to 0.1 Hz. Prior to this measurement, the electrode was held at the potential for 15 min. The resulting impedance spectra were analyzed with ZSimpWin software using suitable equivalent circuit model. All potential values are iR corrected to account for the uncompensated solution resistance between the working and reference electrodes and converted to the RHE scale.

## Acknowledgements:

This work was supported by the University of Southern California (USC) Provost New Strategic Directions for Research Award and The Ershaghi Center for Energy Transition, Viterbi School of Engineering at USC. The authors gratefully acknowledge the use of facilities at the Core Center for Excellence in Nano Imaging (CNI) at USC for the results reported in this manuscript. Use of the Stanford Synchrotron Radiation Lightsource, SLAC National Accelerator Laboratory, is supported by the U.S. Department of Energy, Office of Science, Office of Basic Energy Sciences under Contract No. DE-AC02-76SF00515. We acknowledge Tom Czyszczon-Burton at USC for helping with XPS measurements. We acknowledge helpful discussions with Prof. Shaama Sharada and Prof. Brent Melot of USC.

## Conflict of Interest:

The authors declare no conflict of interest.

## Table of Contents

*We have synthesized a series of novel transition metal-based ternary sulfides by sulfurization of corresponding oxides. These sulfide compounds exhibit excellent electrocatalytic activity towards both halves of the water splitting reaction and outperform their parent oxide materials for both hydrogen and oxygen evolution reactions.*

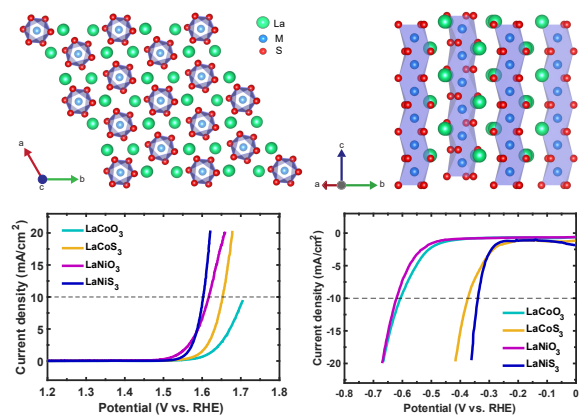

# Supporting Information (SI)

# Bifunctional Noble Metal-free Ternary Chalcogenide Electrocatalysts for Overall Water Splitting

Shantanu Singh[1,5], Ahamed Irshad[2], Germany Diaz De la Cruz[2], Boyang Zhao[1], Billal Zayat[2], Qiaowan Chang[3], Sri Narayan[2], Jayakanth Ravichandran[1,4,5*]

## AUTHOR ADDRESS

1: Mork Family Department of Chemical Engineering and Materials Science, University of Southern California, Los Angeles, California 90089, United States

2: Department of Chemistry, University of Southern California, Los Angeles, California 90089, United States

3: The Gene and Linda Voiland School of Chemical Engineering and Bioengineering, Washington State University, Pullman, Washington 99164, United States

4: Ming Hsieh Department of Electrical and Computer Engineering, University of Southern California, Los Angeles, California 90089, United States

5: Core Center of Excellence in Nano Imaging, University of Southern California, Los Angeles, California, 90089, United States

## Content:

1. **Carbon disulfide Annealing Setup**
2. **Chemical Composition Analysis by SEM-EDS**
3. **XPS and XANES Measurements**
4. **Electrochemical Surface Area (ECSA) Measurements**
5. **LaMS$_3$ Activity towards OER in Acidic Medium**
6. **Effect of Carbon Addition on HER Performance**
7. **Electrochemical Impedance Spectroscopy (EIS) Measurements**

# 1. Carbon disulfide Annealing Setup

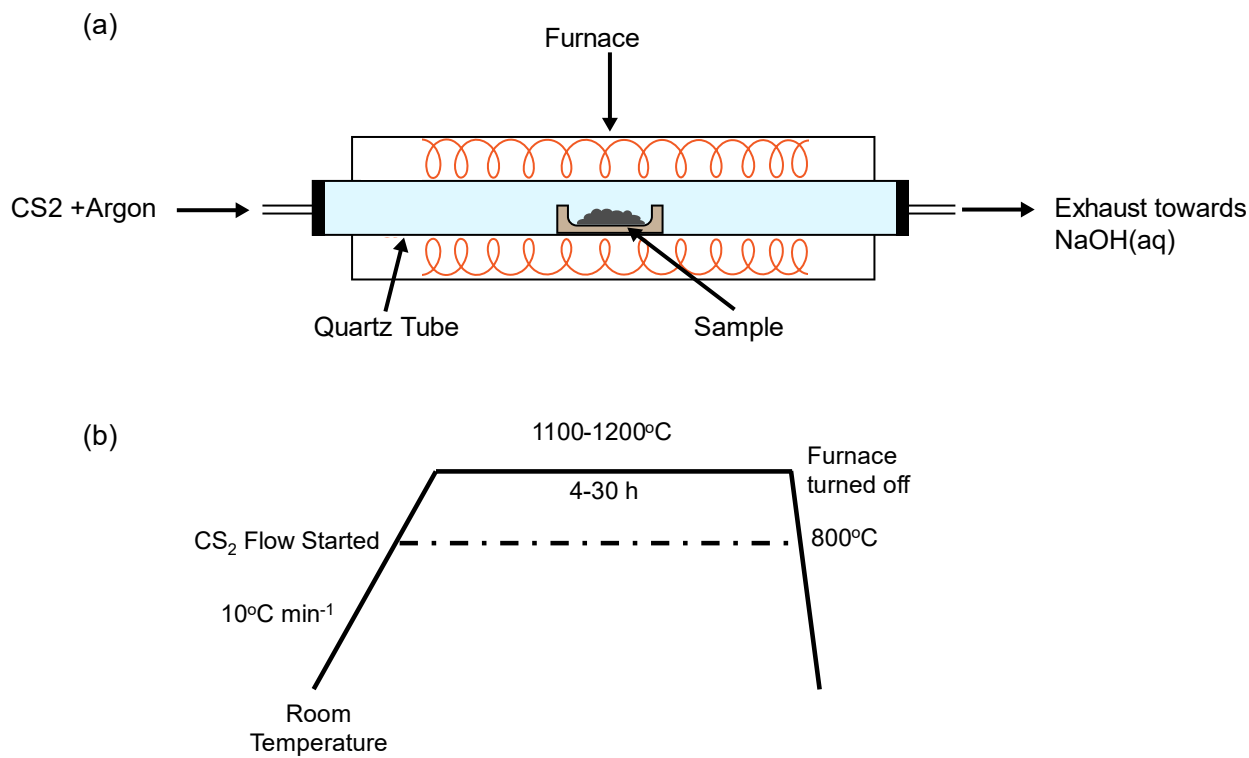

**Figure S1:** **(a)** A schematic representation of the $CS_2$ annealing setup used for the synthesis of $LaMS_3$ powders, and **(b)** the temperature profile used for synthesis.

The schematic for the annealing setup used for the sulfurization of $LaMO_3$ powders to synthesize $LaMS_3$ powders is shown in Figure S1. On the inlet size, a three-way valve is used to switch between Argon and $CS_2$ + Argon gas lines. The $CS_2$ + Argon line has a glass bubbler filled with $CS_2$ liquid, which, when used, bubbles Argon gas through the liquid $CS_2$ column and carries $CS_2$ saturated Argon gas through the quartz tube over the sample. On the exhaust side, the gas is passed through a gas washing bottle filled with 2M aqueous solution of NaOH that neutralizes the excess $CS_2$ vapors.

## 2. Chemical Composition Analysis by SEM-EDS

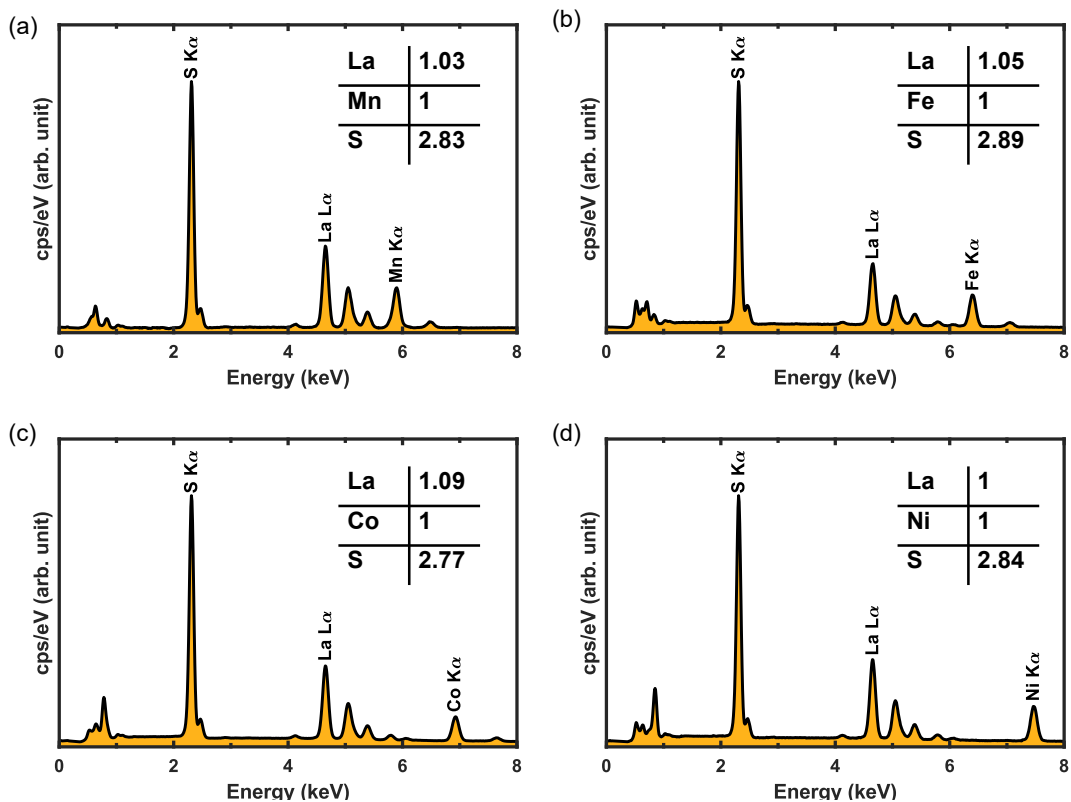

**Figure S2:** EDS Spectrum of **(a)** LaMnS$_3$, **(b)** LaCoS$_3$, **(c)** LaFeS$_3$ and **(d)** LaNiS$_3$ powders. Elemental atomic ratios w.r.t the transition metal elements are shown in the insets.

Energy dispersive spectroscopy (EDS) was used to determine the stoichiometric ratio of elements to determine chemical composition. Measurements were made using 6mm pellets of LaMS$_3$ powders prepared using an MTI YLJ-100E 100T electric hydraulic press and a 900 MPa CIP (cold isostatic press) die set at ~830 MPa for 5 min. The La: M ratios were calibrated using LaMO$_3$ measurements. All the LaMS$_3$ measurements give ~1:1:3 ratios for La: M: S.

# 3. XPS and XANES Measurements

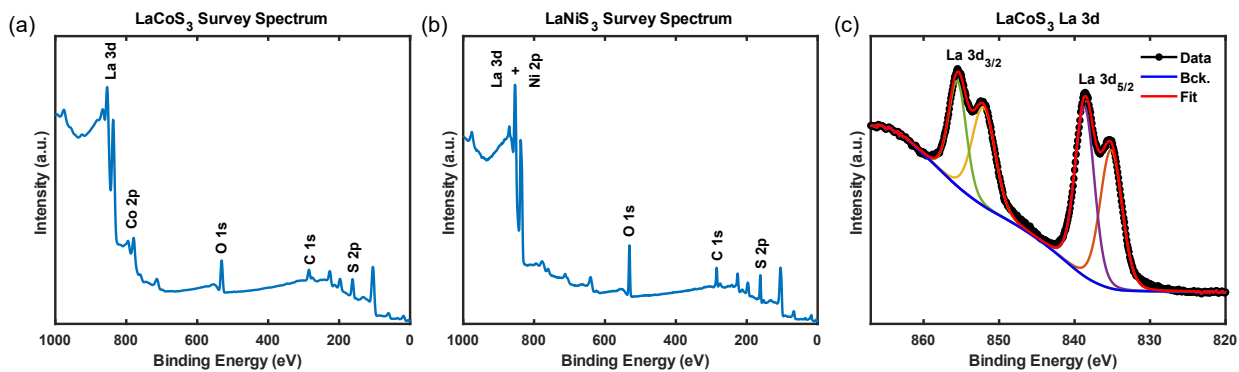

**Figure S3:** Survey spectra for **(a)** LaCoS$_3$, and **(b)** LaNiS$_3$; and **(c)** high resolution XPS spectrum for La 3$d$ peak for LaCoS$_3$.

XPS survey scans were done for LaCoS$_3$ and LaNiS$_3$, and the obtained spectra, indicating the relevant element peaks, are shown in Figure S3(a) and (b), for LaCoS$_3$ and LaNiS$_3$, respectively. The high-resolution XPS spectrum for the La 3$d$ peak in the case of LaCoS$_3$ is shown in Figure S3(c). The fitted peak positions obtained are 835.2 and 838.7 eV for La 3$d_{5/2}$, and 852 and 855.5 eV for La 3$d_{3/2}$. These values are very close to the peak position values obtained by fitting the La 3$d$ + Ni 2$p$ spectra for LaNiS$_3$, where it is 835.2 and 838.6 eV, and 852 and 855.4 eV for La 3$d_{5/2}$ and 3$d_{3/2}$ respectively.

In XANES data processing, the edge energy (E$_0$) was determined by taking the energy of the first large peak from the first derivative of the absorption coefficient vs energy spectra. The E$_0$ values for all the materials reported in Figures 2(c) and (f) are tabulated below.

**Table S1:** XAS Edge Energy (E$_o$) values

| Material | E$_0$ (eV) (Ni-$K$ edge) | Material | E$_0$ (eV) (Co-$K$ edge) |
|---|---|---|---|
| NiS | 8338.4 | CoS | 7714.2 |
| Ni$_3$S$_2$ | 8338 | LaCoS$_3$ | 7714.1 |
| LaNiS$_3$ | 8338.3 | LaCoO$_3$ | 7724.6 |
| LaNiO$_3$ | 8349.1 | | |

# 4. Electrochemical Surface Area (ECSA) Measurements

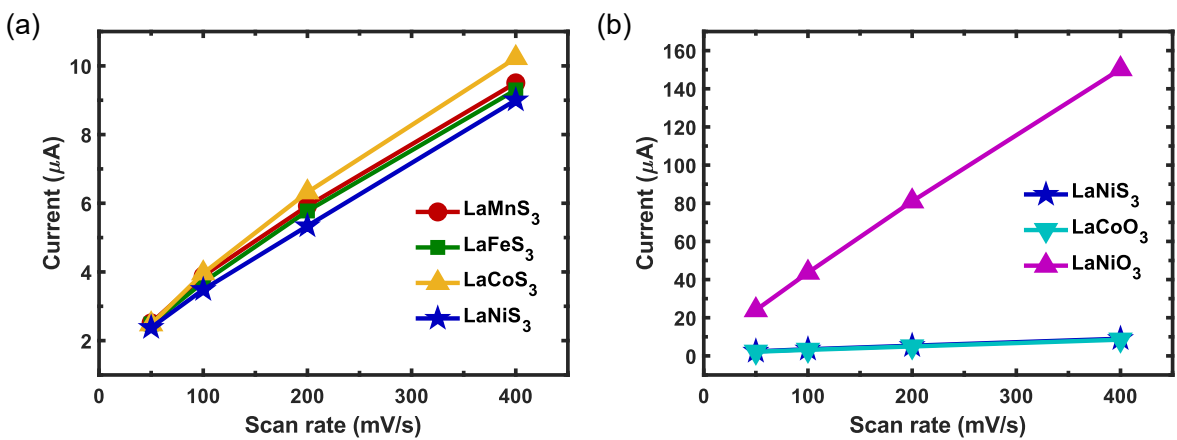

**Figure S4: (a)** Current vs. Scan rate plots for all the sulfides, and **(b)** selected oxides and sulfides in the non-Faradaic region of the cyclic voltammograms in 1 M KOH solution.

ECSA measurements confirm that the trend observed in the activity of LaMS$_3$ materials for OER in 1 M KOH is due to the intrinsic property of the catalysts. The very high surface area of LaNiO$_3$ could be attributed to its nanoparticle morphology. A specific capacitance value of 40 $\mu$F cm$^{-2}$ was used for the ECSA calculation in accordance with electrocatalysis literature in alkaline medium.[1]

**Table S2:** Capacitance, Electrochemical Surface Area (ECSA), and Roughness Factor (RF) of all the catalysts in 1 M KOH electrolyte.

| Catalyst | Capacitance ($\mu$F) | ECSA = Capacitance / 40 (cm$^2$) | RF = ECSA / Geometrical area |
|---|---|---|---|
| LaNiS$_3$ | 19.2 | 0.48 | 2.45 |
| LaCoS$_3$ | 20.5 | 0.51 | 2.60 |
| LaFeS$_3$ | 19.6 | 0.49 | 2.50 |
| LaMnS$_3$ | 20.3 | 0.50 | 2.55 |
| LaNiO$_3$ | 363 | 9.1 | 46.5 |
| LaCoO$_3$ | 18.3 | 0.46 | 2.34 |

## 5. LaMS$_3$ Activity towards OER in Acidic Medium

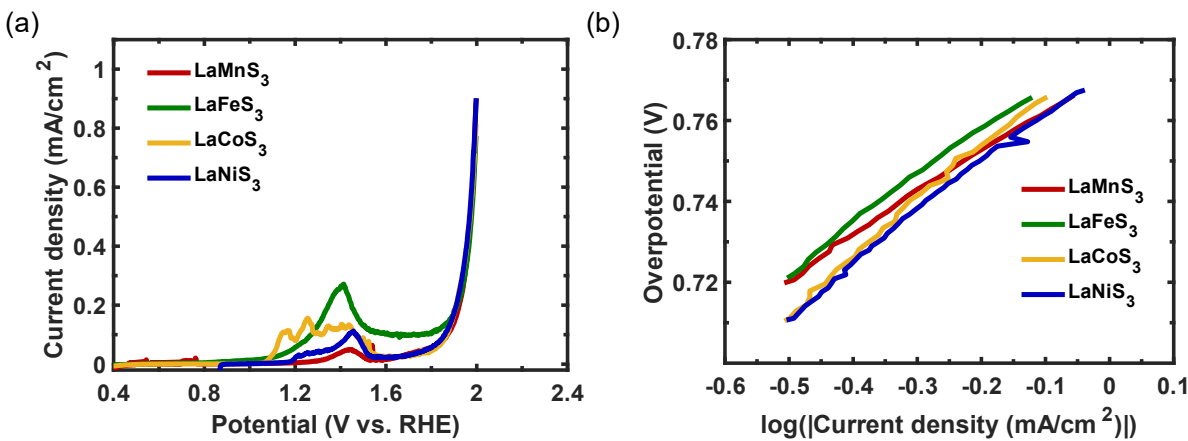

**Figure S5.** OER tests in 0.5 M H$_2$SO$_4$ solution. (a) Linear sweep voltammograms of all the sulfides at 1 mV s$^{-1}$ and 2000 rpm, and (b) corresponding Tafel plots in the linear region. All potential values are iR corrected.

The electrocatalytic OER performance of LaMS$_3$ compounds was tested in 0.5 M H$_2$SO$_4$ solution. Unlike the alkaline medium, no significant activity is observed for any of the catalysts in the acidic medium. The polarization curves also show a broad, intense current peak prior to the onset of OER, indicating the oxidation of the catalyst. The Tafel slopes are notably higher for LaNiS$_3$ and LaCoS$_3$ in acidic electrolytes compared to the alkaline medium, with values of 129 mV dec$^{-1}$ for LaNiS$_3$ and 141 mV dec$^{-1}$ for LaCoS$_3$.

## 6. Effect of Carbon Addition on HER Performance

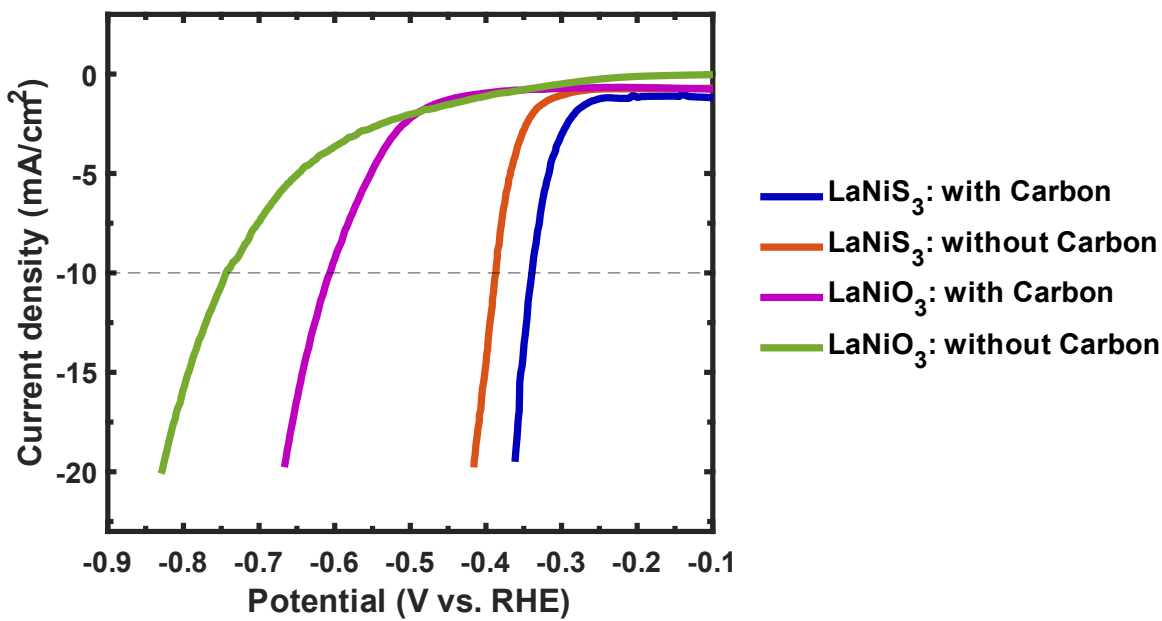

**Figure S6:** Linear sweep voltammograms of $LaNiS_3$ and $LaNiO_3$, with and without SuperP carbon addition in 0.5 M $H_2SO_4$ solution at 1 mV s$^{-1}$ and 2000 rpm. All potential values are iR corrected.

The effect of carbon addition on the HER performance of both oxides and sulfides was studied in 0.5 M $H_2SO_4$. A reduction of activity without carbon in terms of increased overpotential is observed for both $LaNiS_3$ and $LaNiO_3$, but the increase is more significant in the case of $LaNiO_3$ compared to $LaNiS_3$.

# 7. Electrochemical Impedance Spectroscopy (EIS) Measurements

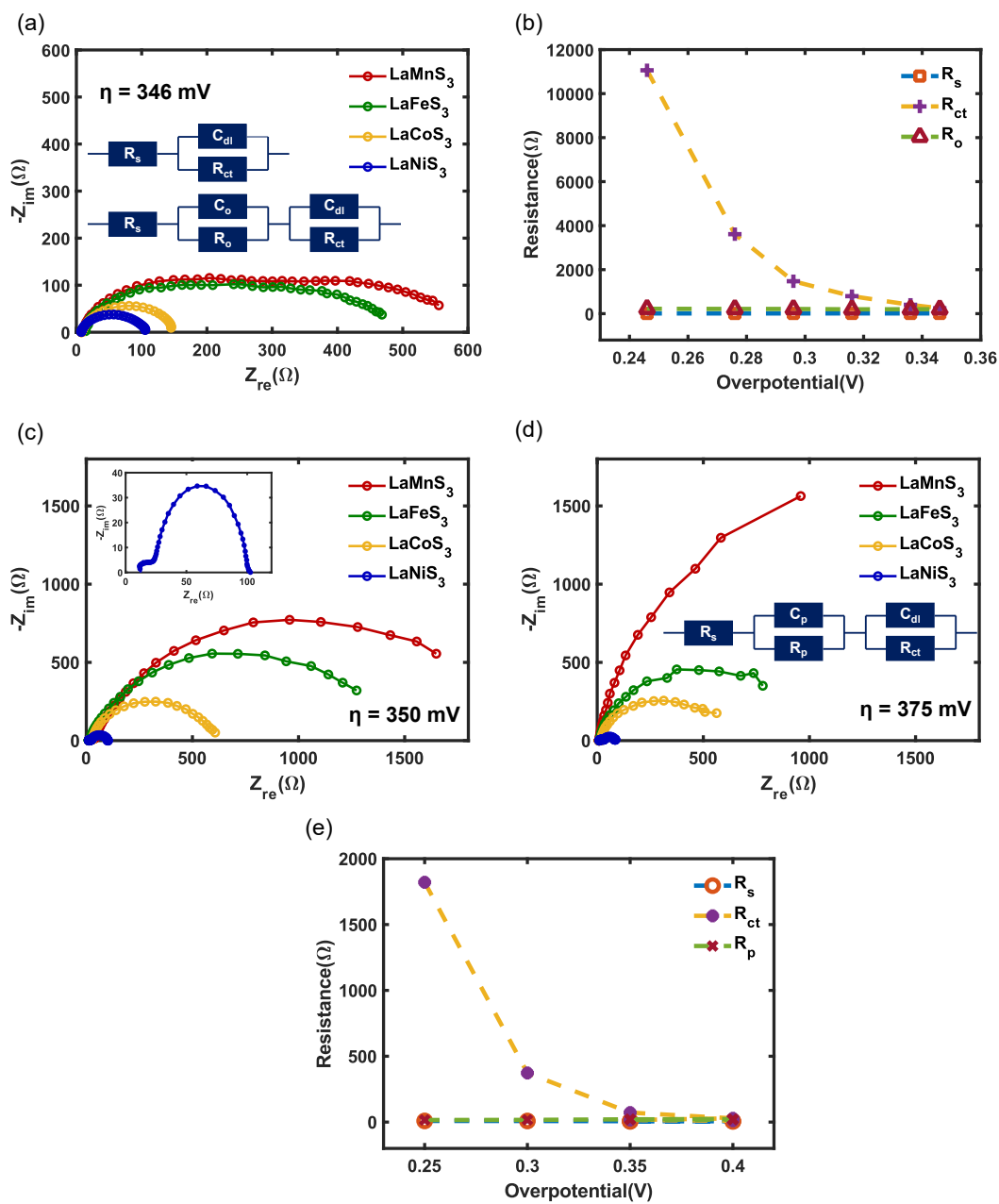

**Figure S7:** EIS measurements for LaMS$_3$ materials. **(a)** EIS during OER in 1 M KOH at 346 mV overpotential. Inset shows the equivalent circuits for LaNiS$_3$ and LaCoS$_3$ (top), and LaFeS$_3$ and LaMnS$_3$ (bottom). **(b)** Variation in $R_s$, $R_o$, and $R_{ct}$ with overpotential for LaFeS$_3$ electrode during EIS for OER in 1 M KOH. **(c)** EIS during HER in 0.5 M H$_2$SO$_4$ at 350 mV overpotential. Inset shows the Nyquist plot of LaNiS$_3$ highlighting the two semicircles. **(d)** EIS during HER in 1 M KOH at 375 mV overpotential. Inset shows the equivalent circuit used for HER in both H$_2$SO$_4$ and KOH. **(e)** Variation in $R_s$, $R_p$ and $R_{ct}$ with overpotential during EIS for LaNiS$_3$ during HER in 0.5 M H$_2$SO$_4$.

Nyquist plots for the LaNiS$_3$ and LaCoS$_3$ catalysts during OER at 346 mV overpotential indicate a semicircle (Figure S7(a)). An equivalent circuit consisting of R$_s$ connected in series with a parallel combination of C$_{dl}$ and R$_{ct}$ (inset in Figure S7(a)) was used to model the spectrum. Here R$_s$ is solution resistance, C$_{dl}$ is constant phase element for the double layer capacitance and R$_{ct}$ is charge transfer resistance for the OER.[2] The LaFeS$_3$ and LaMnS$_3$ catalysts exhibit an additional semicircle at high frequency region (Figure S7(a)). An equivalent circuit with two parallel R-C segments connected in series with the solution resistance was used to model the spectrum (inset in Figure S7(a)). Here, the additional R-C circuit at high frequencies, denoted R$_{oxide}$ (R$_o$) and C$_{oxide}$ (C$_o$), shows minimal dependance on potential, as shown in Figure S7(b), and may be due to anodic surface layer.[3] Such EIS behavior is reported for Fe and Mn based electrolytes in alkaline media.[4-6] EIS on Fe and Mn based electrodes are sensitive to an anodic surface film formation, which is not detected on Ni or Co based electrodes. The LaNiS$_3$ catalyst shows the least charge transfer resistance value of 97 Ω whereas it is 137, 244 and 341 Ω for LaCoS$_3$, LaFeS$_3$ and LaMnS$_3$, respectively. Since the charge transfer resistance is inversely related to rate constant, EIS data suggests highest OER activity for the LaNiS$_3$, consistent with the catalytic activity trend observed in LSV measurements.

In the case of EIS during HER in both acidic and alkaline electrolytes, two semicircles are observed for all the catalysts (Figure S7(c)). The porosity of the electrode surface is responsible for the high-frequency semicircle, which is independent of the applied potential (Figure S7(e)).[2] On the contrary, the low-frequency semicircle shows a strong dependence on the potential, suggesting a charge transfer process at the electrode/ electrolyte interface. An equivalent circuit with two parallel R-C segments connected in series with another resistance is used (inset in Figure S7(d)). At 350 mV overpotential in 0.5 M H$_2$SO$_4$, the charge transfer resistance values for the HER are 72, 560, 1380, and 1953 Ω for $LaNiS_3$, $LaCoS_3$, $LaFeS_3$, and $LaMnS_3$, respectively. Similarly, the charge transfer resistance values of 50, 576, 970, and 4400 Ω are obtained for $LaNiS_3$, $LaCoS_3$, $LaFeS_3$, and $LaMnS_3$, respectively at 375 mV overpotential in 1 M KOH. The EIS results suggest the highest HER activity for $LaNiS_3$, followed by $LaCoS_3$, $LaFeS_3$, and $LaMnS_3$ in the order of decreasing activity.

## SI References: